\documentclass[twocolumn,aps,prb,superscriptaddress,showpacs]{revtex4}
\usepackage{graphicx}
\usepackage{amsmath}
\usepackage{amsfonts}
\usepackage{amssymb}

\begin{document}

\author{Maciej Zwierzycki}
\altaffiliation[Permanent address: ]{Institute of Molecular
Physics, P.A.N., Smoluchowskiego 17, 60-179 Pozna\'n, Poland}
\affiliation{Faculty of Science and Technology and MESA$^{+}$
Research Institute, University of Twente, 7500 AE Enschede, The
Netherlands}
\author{Yaroslav Tserkovnyak}
\affiliation{Lyman Laboratory of Physics, Harvard University,
  Cambridge, Massachusetts 02138, USA}
\author{Paul J. Kelly}
\affiliation{Faculty of Science and Technology and MESA$^{+}$
  Research Institute, University of Twente, 7500 AE Enschede, The Netherlands}
\author{Arne Brataas}
\affiliation{Department of Physics, Norwegian University of Science and Technology,
  N-7491 Trondheim, Norway}
\author{Gerrit E. W. Bauer}
\affiliation{Kavli Institute of NanoScience, Delft University of
Technology,
  2628 CJ Delft, The Netherlands}
\title{First-principles study of
magnetization relaxation enhancement and spin-transfer in thin magnetic films}

\begin{abstract}
The interface-induced magnetization damping of thin ferromagnetic
films in contact with normal-metal layers is calculated from first
principles for clean and disordered Fe/Au and Co/Cu interfaces.
Interference effects arising from coherent scattering turn out to
be very small, consistent with a very small magnetic coherence
length. Because the mixing conductances which govern the spin
transfer are to a good approximation real-valued, the spin pumping
can be described by an increased Gilbert damping factor but an
unmodified gyromagnetic ratio. The results also confirm that the
spin-current induced magnetization torque is an interface effect.
\end{abstract}

\date{\today}
\maketitle


\section{Introduction}

The local magnetization dynamics in a bulk ferromagnet are usually well
described by a phenomenological model formulated in terms of three
parameters: $\mathbf{H}_{\text{eff}}$, an effective magnetic field; $\gamma $%
, a gyromagnetic ratio; and $\alpha $, a Gilbert-damping constant. The field
$\mathbf{H}_{\text{eff}}$ is a sum of contributions from externally applied
fields, crystal anisotropy, shape-dependent dipolar interactions, and
exchange interactions which govern ferromagnetic spin-wave spectral
characteristics. $\gamma $ is the ratio of the total magnetic moment and the
angular momentum of the electrons in the ferromagnet; in 3$d$
transition-metal ferromagnets, such as Fe and Co, it is close to the
free-electron value $\gamma \approx 2\mu _{B}/\hbar $. The Gilbert-damping
constant $\alpha $ parametrizes the viscous damping of an excited
magnetization to the (locally) lowest-energy configuration. Its value
differs considerably for various materials and also depends on the
temperature and on the impurity/defect composition of a given sample. The
motion of the magnetization-direction unit vector $\mathbf{m}$ is determined
by the phenomenological Landau-Lifshitz-Gilbert (LLG) equation:\cite%
{Gilbert:pr55}
\begin{equation}
\frac{d\mathbf{m}}{dt}=-\gamma \mathbf{m\times H}_{\text{eff}}+\alpha
\mathbf{m}\times \frac{d\mathbf{m}}{dt}\,.  \label{llg}
\end{equation}

The magnetization dynamics of small monodomain ferromagnets are well
described by the LLG equation (\ref{llg}) down to the micron scale. New
effects may play a role on the submicron scale, however. The magnetization
dynamics is no longer a highly coherent process because interface and
surface roughness are relatively more important in small samples.
Many-magnon processes can then acquire a sizeable spectral weight \cite%
{Arias:prb99} and are observable as, e.g. an increased line width
of the ferromagnetic resonance (FMR).\cite{Platow:prb98} Another
source of additional FMR broadening is non-local, depending on the
environment into which the ferromagnet is embedded: a
time-dependent ferromagnetic order parameter pumps spin currents
that carry angular momentum (and energy) into adjacent conducting
materials. \cite{Tserkovnyak:prl021,Tserkovnyak:prb022} This
angular-momentum loss, in turn, is equivalent to an additional
damping torque on the magnetization.\cite{Berger:prb96}

The spin-pumping concept for the magnetization dynamics of nanostructures
has far-reaching consequences. It gives rise to an enhanced Gilbert damping
of magnetic films in contact with conducting media,\cite{Tserkovnyak:prl021}
may be employed as an FMR-operated spin battery,\cite{Brataas:prb02}
explains a dynamic exchange coupling in magnetic bilayers,\cite%
{Heinrich:prl03,Tserkovnyak:jap03} as well as a dynamic stiffness
against current-induced magnetization reversal.\cite{Tserkovnyak:prb03}
The analysis of experimental
FMR probes of the magnetization dynamics in single films\cite%
{Tserkovnyak:prb022} and magnetic bilayers\cite{Heinrich:prl03} relied on
phenomenological models of the electronic structure. Here we show how these
assumptions can be relaxed by using instead scattering matrices calculated
from first principles which take into account the detailed atomic and
electronic structure of the materials under study.

An early phenomenological treatment of the non-locality of the
magnetization dynamics in hybrid normal-metal/ferromagnet
(\textit{N{\rm/}F}) structures was given by Silsbee \textit{et
al.}\cite{Silsbee:prb79} Recently, \v{S}im\'{a}nek \textit{et
al.}\cite{Simanek:prb03} pointed out that time-dependent
linear-response theory could be used to calculate the spin flows
generated by a ferromagnet with a time-varying magnetization in
contact with a non-magnetic conductor, as an alternative to the
scattering-theory approach of Tserkovnyak
\textit{et al.}\cite{Tserkovnyak:prl021} In spite of the different starting point, complete
agreement between the two methods was demonstrated \cite{Simanek:prep} for
the simple case of a $\delta $-function magnetic layer embedded in a
free-electron gas. In addition, it was argued in Refs.~%
\onlinecite{Simanek:prb03,Simanek:prep} that the electron-electron
interactions can considerably enhance the spin currents into normal
metals with large Stoner-enhancement factors. The linear-response
framework has also been used to calculate the enhanced Gilbert damping
of finite-thickness ferromagnetic films.\cite{Mills:prb03} It was
argued there that ultra-thin films display oscillatory damping (as a
function of thickness) due to quantum-size effects. In the following,
we show that these quantum-interference effects are greatly
overestimated by the ballistic free-electron band model and do not survive when
realistic transition-metal band structures are used. By calculating
from first principles the scattering matrix entering the spin-pumping
theory,\cite{Tserkovnyak:prl021} we show that quantum-size
oscillations are much smaller than those reported in
Ref.~\onlinecite{Mills:prb03}, especially if even small amounts of
disorder are introduced. We also find that the additional term in the
ferromagnetic equation of motion is of the Gilbert-damping form, with
only a very small correction to the gyromagnetic ratio (the same
conclusion can also be drawn from previous work\cite{Xia:prb02}).
Furthermore, the electron-electron interaction effects discussed by
\v{S}im\'{a}nek\cite{Simanek:prep} are taken into account in the
exchange-correlation potential which we calculate self-consistently
within the local spin density approximation (LSDA) of density-functional
theory. Finally, our results confirm that the spin-current-induced
magnetization torque\cite{Kiselev} is an interface effect, which was
earlier taken for granted \cite{Slon96,Brataas:prl00} and analyzed in
detail in Ref.~\onlinecite{Stiles:prb02}.

This article is organized as follows. The general theory of spin pumping
and its consequences for the dynamics of the precessing ferromagnet are
reviewed in Sec.~\ref{sec:theory}. In Sec.~\ref{sec:first-princ-meth}
we describe the first-principles methods used to obtain the results
presented and discussed in Sec.~\ref{sec:results-discussion}.
A comparison with results based on a free-electron model is made in
an Appendix and conclusions are drawn in Sec.~\ref{sec:conclusions}.

\section{Theory}

\label{sec:theory}

We first consider a ferromagnetic film of thickness $d$ connected to two
perfect non-magnetic reservoirs by two leads which support well-defined
scattering states. The electrons incident on the ferromagnet from a lead are
distributed according to the Fermi-Dirac statistics of the respective
reservoir, whereas the probability that an electron leaving the ferromagnet
returns there with finite spin (or phase) memory is vanishingly small. Such
perfect spin sinks can be realized experimentally by attaching leads to the
ferromagnetic film in the form of point contacts with dimensions smaller
than the electron mean free path. \cite{Myers:sc99} Alternatively, a normal
conductor with a very high spin-flip to momentum scattering-rate ratio (as
could be provided by heavy impurities with large spin-orbit interaction in a
light metal or a heavy metal with phonon or defect scatterers) can serve as
a good spin sink.\cite{Tserkovnyak:prb022}

Coherent motion of the magnetization, whose direction is given by the unit
vector $\mathbf{m}(t)$, leads to the emission of a spin current
\begin{equation}
\mathbf{I}_{s}=\frac{\hbar }{4\pi }\left( \mathrm{Re}A^{\uparrow
\downarrow }\mathbf{m}\times \frac{d\mathbf{m}}{dt}+\mathrm{Im}%
A^{\uparrow \downarrow }\frac{d\mathbf{m}}{dt}\right)
\label{Is}
\end{equation}%
per unit area of the contact into each normal-metal layer,\cite%
{Tserkovnyak:prl021} which we will here assume is then fully absorbed
by the spin sinks (reservoirs).\cite{Tserkovnyak:prb022} The complex
spin-pumping conductance\cite{note1}
\begin{equation}
A^{\uparrow \downarrow }=g_{\uparrow \downarrow }^{r}-g_{\uparrow \downarrow
}^{t}
\end{equation}%
is the difference between the reflection ($g_{\uparrow \downarrow }^{r}$)
and transmission ($g_{\uparrow \downarrow }^{t}$) mixing conductances
(per unit contact area) which
are defined in terms of the spin-dependent reflection and transmission
matrices of the ferromagnetic film as\cite%
{Brataas:prl00,Brataas:spin_echo}
\begin{align}
g_{\uparrow \downarrow }^{r}& =S^{-1}\sum_{mn}\left( \delta
_{mn}-r_{mn}^{\uparrow }r_{mn}^{\downarrow \star }\right) ,
\label{eq:mix_cond} \\
g_{\uparrow \downarrow }^{t}& =S^{-1}\sum_{mn}t_{mn}^{^{\prime }\uparrow
}t_{mn}^{^{\prime }\downarrow \star }.  \label{eq:mix_tran}
\end{align}%
Here, $S$ is the \textit{F{\rm/}N} contact area,
$m$ and $n$ denote scattering states at the Fermi energy of the normal-metal
leads. For spin-pumping into one of the normal-metal layers, $g_{\uparrow
\downarrow }^{r}$
is expressed in terms of the amplitude $r_{mn}^{\sigma }$
for an incoming electron in state $m$ of the normal metal to be reflected at
the interface with the magnetic film into the outgoing state $n$,
while $%
g_{\uparrow \downarrow }^{t}$ is expressed in terms of the amplitude $%
t_{mn}^{^{\prime }\sigma }$ for an incoming
electron from the other \textit{N} layer to be transmitted across
the ferromagnet into the outgoing state $n$.
The total angular-momentum loss of the ferromagnet is given by a sum of
contributions (\ref{Is}) from the two leads, characterized by two
spin-pumping parameters $A^{\uparrow\downarrow}_1$ and $A^{\uparrow\downarrow}_2$.
As explained in Ref.~\onlinecite{Tserkovnyak:prl021},
adding this source of spin angular-momentum current to the right-hand side
of Eq.~(\ref{llg}) leads to a new LLG equation for the monodomain thin film
with saturation magnetization $M_{s}$ embedded in the non-magnetic conducting
medium, with the modified constants $\alpha _{\text{eff}}$ and $\gamma _{%
\text{eff}}$
\begin{align}
\frac{1}{\gamma _{\text{eff}}}& =\frac{1}{\gamma }\left[ 1-\frac{\hbar
\gamma }{4\pi M_{s}d}\mathrm{Im}\left(A_1^{\uparrow\downarrow}+A_2^{\uparrow\downarrow}\right)\right] ,
\label{geff} \\
\alpha _{\text{eff}}& =\frac{\gamma _{\text{eff}}}{\gamma }\left[ \alpha +%
\frac{\hbar \gamma }{4\pi M_{s}d}\mathrm{Re}\left(A_1^{\uparrow\downarrow}+A_2^{\uparrow\downarrow}\right)\right] .
\label{aeff}
\end{align}%
It can be easily shown\cite{Tserkovnyak:prl021} that the real part of
$A^{\uparrow \downarrow }$ is always non-negative so that the correction
to the damping is always positive.
The reader is referred to Sec.~\ref{sec:results-discussion} for a
discussion of the absolute and relative values of
$g_{\uparrow \downarrow }^{r}$ and $g_{\uparrow \downarrow }^{t}$.
Anticipating these results, we note here that in typical
situations $g_{\uparrow \downarrow }^{t}$ and
$\mathrm{Im}g_{\uparrow \downarrow }^{r}$
(and thus $\mathrm{Im}A^{\uparrow \downarrow }$) are negligible so that
the only effect of the spin pumping is to make an additional contribution
to the Gilbert-damping parameter. We
shall therefore assume for the rest of the current section that
$g_{\uparrow\downarrow }^{t}\ll g_{\uparrow \downarrow }^{r}$ with the
latter quantity being essentially an interface property.

Eq.~(\ref{Is}) was derived for an \textit{N{\rm/}F{\rm/}N} structure
connected to perfectly equilibrated reservoirs.\cite{Tserkovnyak:prl021,Tserkovnyak:prb022}
By using this geometry, the finiteness of the Sharvin conductances is
automatically included.\cite{Datta:book} To apply calculated mixing
conductances to the discussion of spin transport in diffuse systems which
are not ideal spin sinks, the ``bare'' conductance \eqref{eq:mix_cond}
has to be corrected\cite{note2} for the corresponding ``spurious''
Sharvin resistance as discussed in Ref.~\onlinecite{Bauer:prb03}.
Additionally, a non-vanishing backflow and reabsorption
of the spins emitted by the ferromagnet has to be taken into
account. The latter can be achieved by considering the diffusion
equation for the spin accumulation in the normal lead with
Eq.~\eqref{Is} providing the boundary condition
(see Ref.~\onlinecite{Tserkovnyak:prb022}).
This leads to an effective (complex) conductance
${\tilde{A}^{\uparrow \downarrow }}$ (for either interface) entering
equations (\ref{geff}) and (\ref{aeff}) where
\begin{equation}
\frac{1}{\tilde{A}^{\uparrow \downarrow }}=\frac{1}{g_{\uparrow \downarrow
}^{r}}-\frac{1}{2g_{N}^{\text{Sh}}}+\frac{2e^{2}}{h}\cdot \frac{R_{\text{SD}}%
}{\tanh (L/\lambda _{\text{SD}})}\, , \label{tA}
\end{equation}%
$g_{N}^{\text{Sh}}$ is the Sharvin conductance of the normal-metal
layer, given by the number of the transverse channels per spin and
unit area of the interface;\cite{Bauer:prb03} $R_{\text{SD}}=\lambda
_{\text{SD}}/\sigma $ is the unit-area resistance of the normal-metal
film with conductivity $\sigma /2$ (per spin) and thickness $\lambda
_{\text{SD}}$, the spin-diffusion length; $L$ is the actual thickness
of the normal-metal layer. The last term on the right-hand side of
Eq.~(\ref{tA}) accounts for impurity, defect, or phonon scattering in
the normal metal. (Scattering in the ferromagnet on length scales
longer than the transverse spin-coherence length does not modify the
result.) When spin-flip scattering in the \textit{N} layer vanishes,
$\lambda _{\text{SD}}\to \infty $, $\tilde{A}^{\uparrow \downarrow
}\to 0$ (\emph{i.e} the backflow spin-current completely cancels the
pumping effect) and the magnetization dynamics is not modified at all.

A similar analysis can be applied to magnetic damping in more complex
multilayer systems.\cite{Tserkovnyak:prb022,Heinrich:prl03}
For example, in an \textit{F{\rm/}N{\rm/}F} structure the presence of
two ferromagnetic layers can make damping possible for each individual
layer even in the absence of spin-flip relaxation in the system.
In this case, each magnetic layer acts as the sink for the spin current
pumped by the other layer. If the structure is weakly excited from a
collinear equilibrium state, and the individual ferromagnetic
resonances are well separated, then a different effective conductance
enters Eqs.~(\ref{geff}) and (\ref{aeff}).
Instead of the sum $A_1^{\uparrow\downarrow}+A_2^{\uparrow\downarrow}$,
for the two magnetic films, the quantity
${\tilde{A}_{F\mathrm{/}N\mathrm{/}F}^{\uparrow \downarrow }}$ with
\begin{equation}
\frac{1}{\tilde{A}_{F\mathrm{/}N\mathrm{/}F}^{\uparrow \downarrow }}
= \frac{1}{g_{\uparrow \downarrow }^{1r}}+\frac{1}{g_{\uparrow \downarrow }^{2r}}
- \frac{1}{g_{N}^{\text{Sh}}}
+ \frac{2e^{2}}{h} \cdot \frac{L}{\sigma } \,  \label{Af}
\end{equation}
defined for the globally diffuse system should be
used, where $g^{ir}_{\uparrow\downarrow}$ is the
mixing conductance for the $i$-th \textit{F\textrm{/}N} interface.
Eq.~(\ref{Af}) can be intuitively interpreted in terms of resistances in
series:
in order to be absorbed, the spin current must be pumped through the
first \textit{F\textrm{/}N} interface ($g^{1r}_{\uparrow\downarrow}$
renormalized by $2 g^{\text{Sh}}_N$), propagate across the normal layer
($L/\sigma$ term) and enter the second ferromagnet through the other
interface ($g^{2r}_{\uparrow\downarrow}$ renormalized by $2 g^{\text{Sh}}_N$).
The formula for ${\tilde{A}_{F\mathrm{/}N\mathrm{/}F}^{\uparrow \downarrow }}$
can be straightforwardly derived using the spin-diffusion approach of
Ref.~\onlinecite{Tserkovnyak:prb022}. It is worthwhile pointing out
that it remains correct for non-diffusive normal metal spacers
($\sigma\rightarrow\infty$) if the interface disorder is sufficient to
suppress any quantum-size effects (see Ref. \onlinecite{Bauer:prb03}).



The effect of spin-dependent scattering on the time evolution of the
magnetic order parameter is therefore mostly governed by three parameters: the
reflection and transmission mixing conductances of the ferromagnetic
layer, $%
g_{\uparrow \downarrow }^{r}$ and $g_{\uparrow \downarrow }^{t}$, and the
Sharvin conductance of the normal metal, $g_{N}^{\text{Sh}}$. We noted before%
\cite{Tserkovnyak:prl021} that these quantities are in principle accessible
to \textit{ab initio} electronic-structure calculations.\cite{Xia:prb02,Xia:prb01,Xia:prl02}
In the following we demonstrate this by studying two representative \textit{N%
{\rm/}F} material combinations: Au/Fe(001) and Cu/Co(111), the former
routinely used by the Simon-Fraser group\cite%
{Urban:prl01,Heinrich:jap02,Heinrich} and the latter by the Cornell group.%
\cite{Myers:sc99,Katine:prl00}

\begin{table}[t]
\begin{center}
\begin{ruledtabular}
\begin{tabular} [c]{lrrrr}
\multicolumn{1}{c}{\textit{N{\rm/}F}} & \multicolumn{2}{c}{Au/Fe}
                                 & \multicolumn{2}{c}{Cu/Co} \\
\hline
Layer&           clean  & dirty  &    clean   &    dirty  \\
\hline
$m_N$(bulk)     & 0.000 &  0.000 & 0.000 & 0.000 \\
$m_N$(int-4)    & 0.000 &  0.000 & 0.001 & 0.000 \\
$m_N$(int-3)    & 0.001 & -0.003 &-0.000 &-0.003 \\
$m_N$(int-2)    &-0.002 &  0.010 &-0.004 &-0.003 \\
$m_N$(int-1)    & 0.064 &  0.026 & 0.006 & 0.010 \\
$m_F$(int-1)    &   -   &  2.742 &  -    & 1.410 \\
\hline
$m_N$(int+1)    &   -   &  0.128 &  -    & 0.036 \\
$m_F$(int+1)    & 2.687 &  2.691 & 1.545 & 1.540 \\
$m_F$(int+2)    & 2.336 &  2.396 & 1.635 & 1.596 \\
$m_F$(int+3)    & 2.325 &  2.363 & 1.621 & 1.627 \\
$m_F$(int+4)    & 2.238 &  2.282 & 1.627 & 1.624 \\
$m_F$(bulk)     & 2.210 &  2.210 & 1.622 & 1.622 \\
\end{tabular}
\end{ruledtabular}
\end{center}
\caption{Layer-resolved magnetic moments in Bohr magnetons for single
\textit{N{\rm/}F} interfaces (\textit{N}=Au, Cu; \textit{F}=Fe, Co).}
\label{tableI}
\end{table}

\section{First-principles method}

\label{sec:first-princ-meth}

Parameter-free calculations of transmission and reflection
coefficients were performed using the local spin density approximation
(LSDA) of density-functional theory (DFT) in a two-step procedure. In
the first step, the self-consistent electronic structure (spin
densities and potentials) of the system was determined using the layer
TB-LMTO (tight-binding linear muffin-tin orbital) surface Green's
function (SGF) method in the atomic-sphere approximation
(ASA).\cite{Turek} The exchange-correlation potential in the
Perdew-Zunger\cite{Perdew:prb81} parametrization was used.
The atomic-sphere (AS)
potentials of 4 monolayers on either side of the magnetic layer (or
interface) were iterated to self-consistency while the potentials of
more distant layers were held fixed at their bulk values. Because both
of the systems we consider, Au/Fe(001) and Cu/Co(111), are nearly
ideally lattice matched, common lattice constants were assumed for
both metals of a given structure: $a_{\mathrm{Cu/Co}}=3.549$ \AA\ and
$a_{\mathrm{Au/Fe}}=\sqrt{2}\times 2.866$ =4.053 \AA. In the second
step, the AS potentials serve as inputs to calculate scattering
coefficients using a recently-developed scheme based on
TB-MTOs.\cite{Xia:prb01,Zwierzycki:prb03,Xia:tbp}
Disorder is modeled by allowing a number of
interface layers to consist of \textit{N}$_{x}$\textit{F}$_{1-x}$ alloy
which is modeled using
repeated lateral supercells. Because a minimal basis set of $s$, $p$ and $d$
orbitals is used, we are able to treat lateral supercells containing as many
as 200 atoms in which the two types of atoms are distributed at random in
the appropriate concentration. For disordered interfaces, the AS potentials
were calculated self-consistently using the layer CPA approximation in which
each layer can have a different alloy composition.\cite{Turek}

Little is known from experiment about the atomic structure of metallic
interfaces. We model \textquotedblleft dirty\textquotedblright\
interfaces with one (for \textit{N{\rm/}F{\rm/}N} systems) or two (for
single \textit{N{\rm/}F} interfaces) atomic layers of a 50~\%-50~\%
alloy. Such a model is probably reasonable for
$\mathrm{Cu_{fcc}/Co_{fcc}}$ because of the nearly perfect lattice
match and structural compatibility. The situation is however more
complicated for $\mathrm{Au_{fcc}/Fe_{bcc}}$ because of the large
difference in AS sizes for Au and Fe with Wigner-Seitz radii of 2.99
and 2.67 Bohr atomic units, respectively. We have assumed here that
the disorder is only substitutional and that the diffused atoms occupy
the AS of the same size as that of the host element. In the Au/Fe/Au
case, where the alloy is only 1 atomic monolayer (ML) thick, we assume
that the Fe atoms diffuse into Au.  While the validity of this model
can be questioned, the insensitivity of the final results to the
details of the disorder (e.g. one versus two monolayers of alloy)
indicate that this is not a critical issue. The layer-resolved
magnetic moments for single interfaces are given in Table~\ref{tableI}.
They agree well with values reported previously in the
literature.\cite{Wang:jmmm98,Samant:prl94,Krompiewski:jmmm95}

The two-dimensional Brillouin zone (2D~BZ) summation required to calculate
the mixing conductances using Eqs. (\ref{eq:mix_cond}) and (\ref{eq:mix_tran}%
) was performed using $k_{||}$-mesh densities corresponding to $10^{4}$ points
in the 2D BZ of a 1$\times $1 interface unit cell. The uncertainties resulting
from this BZ summation and from impurity ensemble averaging are of the order
of a few times $10^{12}~\Omega ^{-1}$m$^{-2}$, which is smaller than the
size of the symbols used in the figures.

\begin{figure}[tb]
\includegraphics[width=0.365\textheight,clip=]{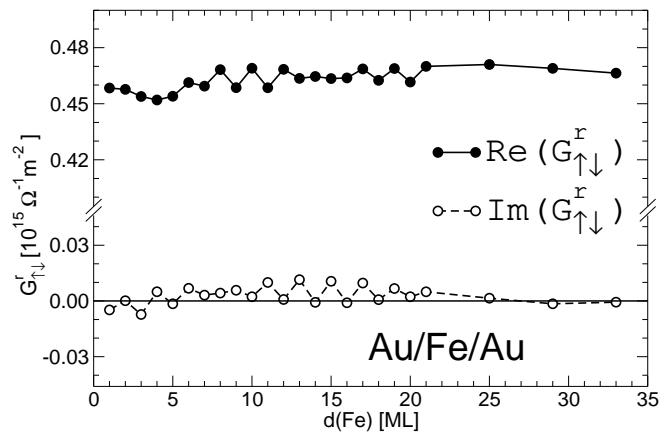}
\caption{Reflection spin-mixing conductance (per unit area) of a
Au/Fe/Au(001) trilayer with perfect interfaces as a function of the thickness $d$ of
the Fe layer. In this and subsequent plots, mixing conductances expressed in
terms of number of conduction channels per unit area are converted to $%
\Omega^{-1}$m$^{-2}$ using the conductance quantum $e^2/h$, i.e.
$G_{\uparrow\downarrow}=(e^2/h)g_{\uparrow\downarrow}$.}
\label{gafc}
\end{figure}

\begin{figure}[tb]
\includegraphics[width=0.365\textheight,clip=]{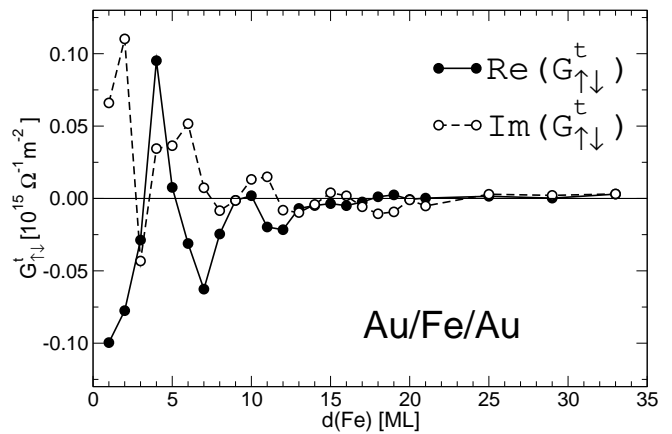}
\caption{Transmission spin-mixing conductance of a Au/Fe/Au (001) trilayer
with perfect interfaces as a function of the thickness $d$ of the Fe layer.}
\label{tafc}
\end{figure}

\begin{figure}[t]
\includegraphics[width=0.365\textheight,clip=]{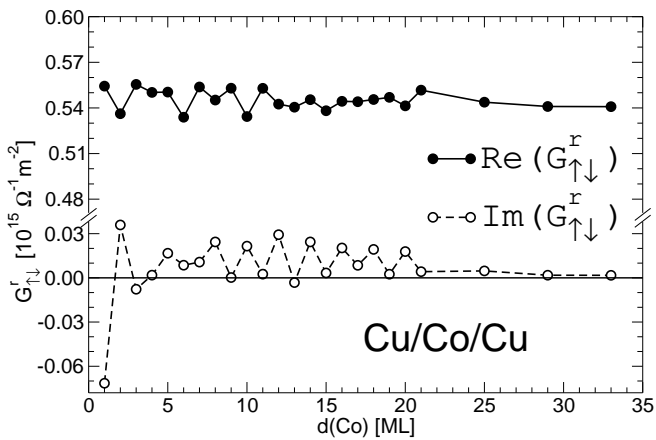}
\caption{Reflection spin-mixing conductance of a Cu/Co/Cu (111) trilayer
with perfect interfaces as a function of the thickness $d$ of the Co layer.}
\label{gccc}
\end{figure}

\begin{figure}[t]
\includegraphics[width=0.365\textheight,clip=]{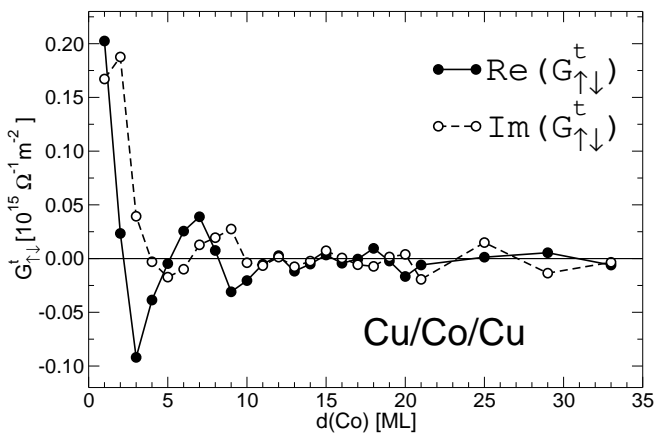}
\caption{Transmission spin-mixing conductance of a Cu/Co/Cu (111) trilayer
  with perfect interfaces as a function of the thickness $d$ of the Co
  layer.}
\label{tccc}
\end{figure}

\begin{figure*}[t]
\centering
\includegraphics[width=0.8\linewidth,clip=]{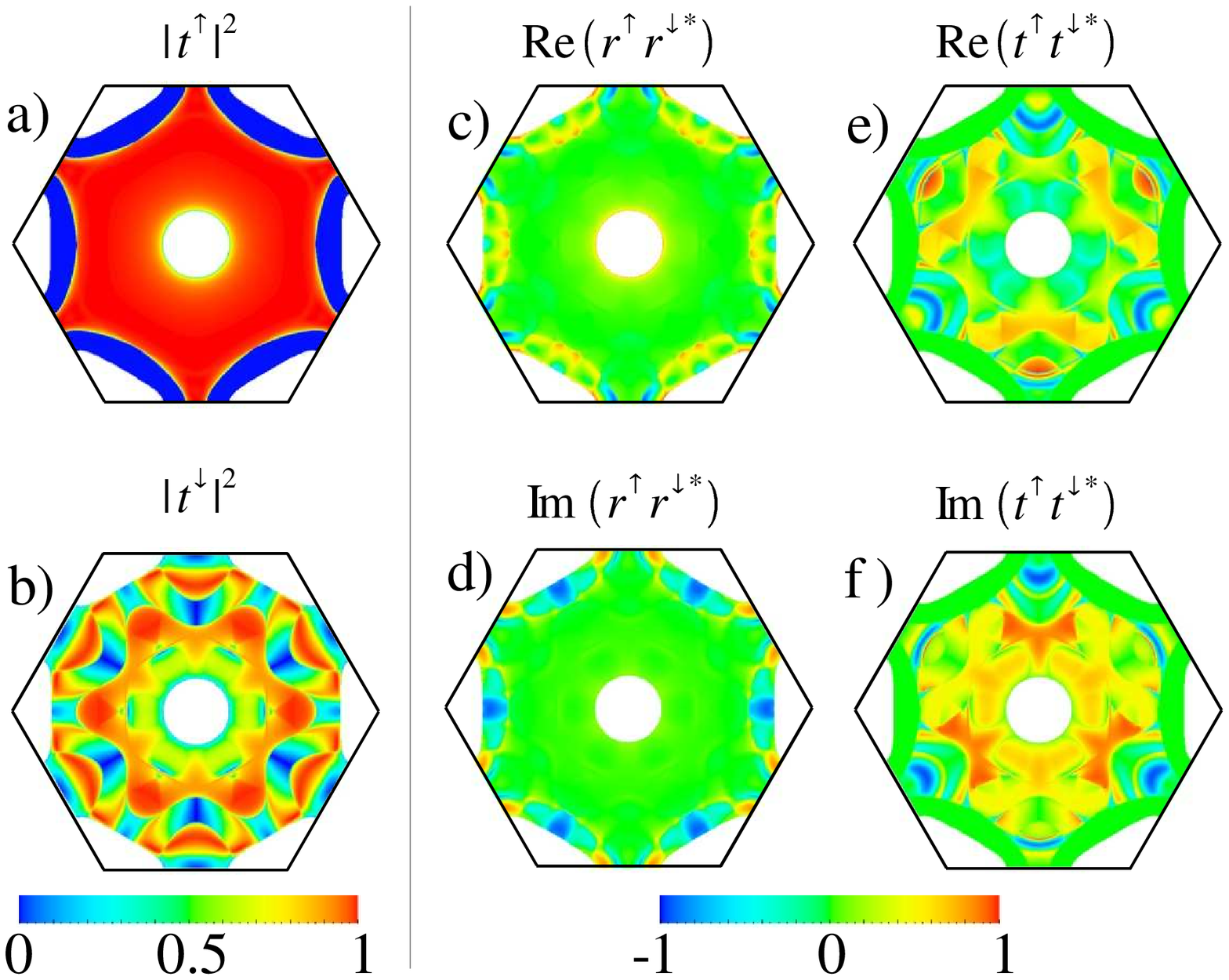}
\caption{Plotted within the first Brillouin zone for the
Cu/Co(111) interface: transmission probability for (a) majority-
and (b) minority-spins. (c) Real and (d) imaginary parts of
$r_{N\to N}^{\uparrow }r_{N\to N}^{\downarrow \star }$. (e) Real
and (f) imaginary parts of $t_{\mathrm{int}}^{\uparrow
}t_{\mathrm{int}}^{\downarrow
\star }$ where $t^{\sigma}_{\mathrm{int}}=t^{\sigma} %
_{F\to N}\cdot t^{\sigma}_{N\to F}$ as discussed in
the text. Note the different scales for panels (a), (b) and for (c) - (f).}
\label{fig:2d_plots}
\end{figure*}

\section{Results and discussion}

\label{sec:results-discussion}
Figures \ref{gafc} to \ref{tccc} show how $%
G_{\uparrow \downarrow }^{r}=(e^{2}/h)g_{\uparrow \downarrow }^{r}$ and $%
G_{\uparrow \downarrow }^{t}=(e^{2}/h)g_{\uparrow \downarrow }^{t}$ depend
on the thickness $d$ of the magnetic layer (measured in atomic layers) for
specular (${\vec{k}}_{||}$-preserving) Au/Fe/Au(001) and Cu/Co/Cu(111)
systems. Both quantities exhibit oscillatory behavior with, however,
noticeably different periods and amplitudes. The values of both $G_{\uparrow
\downarrow }^{r}$ and $G_{\uparrow \downarrow }^{t}$ are determined by two
factors: the matching of the normal metal and ferromagnetic
metal states at the interface (described by the scattering coefficients of the single
interface) and the phases accumulated by electrons on their passage through
the magnetic layer (quantum-size effect). The first factor determines the
amplitudes of the oscillations and (for $G_{\uparrow \downarrow }^{r}$) the
asymptotic values, while the second is responsible for the observed
periodicity. In order to better understand this, it is instructive to
interpret the transmission and reflection coefficients of the finite-size
magnetic layer in terms of multiple scattering at the interfaces. We first
note that both Cu and Au have only one left- and one right-going state at
the Fermi level for each value of $\vec{k}_{||}$ and spin so that the
summations in Eqs.~(\ref{eq:mix_cond}) and (\ref{eq:mix_tran}) reduce to
integrations over the 2D BZ involving the complex-valued functions $%
r^{\sigma }(\vec{k}_{||})$ and $t^{\sigma }(\vec{k}_{||})$.
Retaining only lowest-order thickness-dependent terms, dropping explicit
reference to $\vec{k}_{||}$ and to the primes on $t^{^{\prime }}$, we
then have
\begin{align}
t^{\sigma }& \approx t_{F\to N}^{\sigma }\Lambda ^{\sigma
}t_{N\to F}^{\sigma }  \label{eq:msc_t} \\
r^{\sigma }& \approx r_{N\to N}^{\sigma }+t_{F\to N}^{\sigma
}\Lambda ^{\sigma }r_{F\to F}^{\sigma }\Lambda ^{\sigma
}t_{N\to F}^{\sigma }  \label{eq:msc_refl}
\end{align}%
where $t_{N\to F}^{\sigma }=(t_{1}^{\sigma },\ldots ,t_{n}^{\sigma
})^{T}$ is a vector of transmission coefficients between a single
propagating state in the normal metal and a set of states in the
ferromagnet, $\Lambda ^{\sigma }$ is a diagonal matrix of phase factors $%
e^{ik_{j\perp }^{\sigma }d}$ ($j$ is an index of the states in the
ferromagnet), $r_{N\to N}^{\sigma }$ is a scalar reflection
coefficient for states incoming from the normal metal and $r_{F\to
F}^{\sigma }$ is a square matrix
describing reflection on the ferromagnetic
side. The set of states in the ferromagnet consists of both propagating and
evanescent states. The contribution of the latter decreases exponentially
with the thickness of the layer.

Concentrating first on the thickness dependence of $g_{\uparrow \downarrow
}^{t}$, we notice that,
in view of Eq.~(\ref{eq:msc_t}), the summation in
Eq.~(\ref{eq:mix_tran}) is carried out over terms containing phase factors $%
e^{i(k_{i\perp }^{\uparrow }-k_{j\perp }^{\downarrow })d}$.
Because of the large differences between majority and minority Fermi surfaces of the
ferromagnet, this typically leads to rapidly oscillating terms which mostly
cancel out on summing over $\vec{k}_{||}$. It can be argued\cite%
{Stiles:prb02} in the spirit of the theory of interlayer exchange coupling%
\cite{Bruno:prb95} that the only long-range contributions originate from the
vicinity of points for which $\nabla _{k_{||}}(k_{i\perp }^{\uparrow
}-k_{j\perp }^{\downarrow })=0$, corresponding to the stationary phase of
the summand in Eq.~(\ref{eq:mix_tran}). These contributions will then
exhibit damped oscillations around zero value as seen in Figs.~\ref{tafc}
and \ref{tccc}.

Turning to $g_{\uparrow \downarrow }^{r}$, we find on substituting Eq.~(\ref%
{eq:msc_refl}) into Eq.~(\ref{eq:mix_cond}) that there are two
thickness-independent contributions. The first comes from summing
the $%
\delta _{nm}$ term in Eq.~(\ref{eq:mix_cond}) and is nothing other than the
number of states in the normal metal (i.e. the Sharvin conductance). The
second comes from the $r_{N\to N}^{\uparrow }r_{N\to
N}^{\downarrow \ast }$ term and provides an interface-specific correction to
the first. Superimposed on these two is the contribution from the
thickness-dependent terms which, to lowest order, contain phase factors
 $%
e^{i(k_{i\perp }^{\sigma }+k_{j\perp }^{\sigma })d}$ and $e^{-i(k_{i\perp
}^{\sigma }+k_{j\perp }^{\sigma })d}$. Just as in the case of $g_{\uparrow
\downarrow }^{t}$, one can argue that
the integral over these terms will
have oscillatory character. However, the oscillations will have different
periods and occur around the constant value set by the first two
contributions. It is clear that the value approached asymptotically by $%
g_{\uparrow \downarrow }^{r}$
is simply the reflection mixing conductance evaluated for a single interface.

The period and damping of oscillations of $g^r_{\uparrow\downarrow}$ and
$g^t_{\uparrow\downarrow}$ as a function of the magnetic-layer
thickness $d$ clearly depends (through the $\Lambda^\sigma$) on the
electronic structure of the internal part of the magnetic layer, which
for metallic systems is practically identical to that of the bulk
material. The amplitudes, on the other hand, are related to the
interfacial scattering coefficients introduced in
Eqs.~(\ref{eq:msc_t}) and (\ref{eq:msc_refl}). Analyzing the
scattering properties of the single interface enables us in the
following to understand why the amplitudes of oscillation of
$g^t_{\uparrow\downarrow}$ are substantially larger than those of
$g^r_{\uparrow\downarrow}$ for the two systems considered. We begin by
noting that the transmission probability for states in the
majority-spin channel assumes values close to one over large areas of
the Brillouin zone for both Cu/Co and Au/Fe, as illustrated in
Fig.~\ref{fig:2d_plots}(a) for the Cu/Co(111) interface. For Cu/Co,
this results from the close similarity of the corresponding Cu and Co
electronic structures. The situation is more complicated for Au/Fe
because the majority-spin Fermi surface of Fe consists of several
sheets, unlike that of Au. However, one of these sheets is made up of
states which match well with the states in Au. In the minority-spin
channel, on the other hand, the
transmission probability varies between 0 and 1; see Fig.~\ref{fig:2d_plots}%
(b). The maximum size of the (absolute value of the) ``spin-mixing''
products of Eqs.~\eqref{eq:mix_cond} and \eqref{eq:mix_tran} are therefore
determined mostly by the majority-spin scattering coefficients while the
modulation, as a function of $\vec{k}_{||}$, is governed by the
corresponding minority-spin coefficients.

The small reflectivity for the majority-spin states has a direct consequence
for the values of the mixing conductances. In the case of $g_{\uparrow
\downarrow }^{r}$, the second term under the sum in Eq.~(\ref{eq:mix_cond})
will typically have a negligible magnitude. This follows directly from $%
r_{N\to N}^{\uparrow }\approx 0$  and Eq.~(\ref{eq:msc_refl}%
) and is illustrated in Fig.~\ref{fig:2d_plots} (c) and (d) for the $%
r_{N\to N}^{\uparrow }r_{N\to N}^{\downarrow \ast }$ term.
As we can see, the only non-zero contributions in this case come from the
outer regions of the Brillouin zone, where states from the normal metal are
perfectly reflected because of the absence of propagating majority-spin
states in the ferromagnet. Independently varying phases (as a function of $%
\vec{k}_{||}$) for \textquotedblleft up\textquotedblright\ and
\textquotedblleft down\textquotedblright\ reflection coefficients lead, in
the course of integration over $\vec{k}_{||}$, to additional cancellation of
already small contributions. The final outcome is that the values of $%
g_{\uparrow \downarrow }^{r}$ are determined mostly by the first term in the
Eq.~(\ref{eq:mix_cond}), i.e. the Sharvin conductance of the lead.

Because the interface transmission in the majority-spin channel is
uniformly large almost everywhere in the Brillouin zone, the transmission
through the magnetic layer also remains large for arbitrary thicknesses
and its magnitude (but not its phase) is only weakly modulated by the
multiple scattering within the layer. The magnitude of the
$t^{\uparrow}t^{\downarrow \ast }$ product is then modulated
mostly by the variation of the transmission in the minority-spin channel,
as a function of $\vec{k}_{||} $. To demonstrate the effect of the interface
scattering on $g_{\uparrow \downarrow }^{t}$,
values of the product
$t_{\mathrm{int}}^{\uparrow }t_{\mathrm{int}}^{\downarrow \ast }$
are shown in Figs.~\ref{fig:2d_plots}(e) and (f) for
a Cu/Co (111) interface. Here, $t_{\mathrm{int}}^{\uparrow }$ is defined
as the scalar product of the interface transmission vectors:
$t_{\mathrm{int}}^{\sigma }=t_{F\to N}^{\sigma}\cdot t_{N\to F}^{\sigma }$.
As one can see, the values assumed by the real and imaginary part of this
product vary strongly throughout the Brillouin zone. Unlike the case of
$g_{\uparrow \downarrow }^{r}$, however,
the values span the entire range from -1 to +1. An imbalance of positive and
negative contributions is therefore more likely to produce a sizeable
integrated value.
The complex values of $t^{\uparrow }t^{\downarrow \ast }$ are further
modified by thickness- and $\vec{k}_{||}$-dependent phase factors
discussed above, which leads to the oscillatory damping seen in
Figs.~\ref{tafc} and \ref{tccc}.
We compare the magnitude and damping of these oscillations with those
derived from a free-electron model in an Appendix.

\begin{figure}[tb]
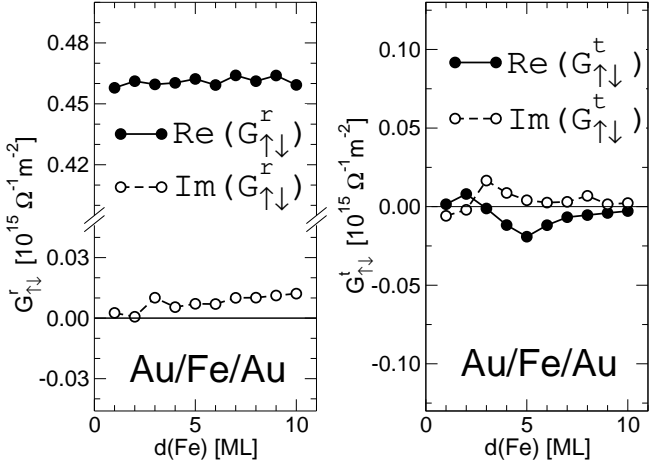

\includegraphics[height=0.26\textheight,clip=]{gmix_aufe_dis.eps} %
\includegraphics[height=0.26\textheight,clip=]{tmix_aufe_dis.eps}
\caption{Spin-mixing conductances of a Au/Fe/Au (001) trilayer with
disordered interfaces as a function of the thickness $d$ of the Fe layer.}
\label{gafd}
\end{figure}

\begin{figure}[t]
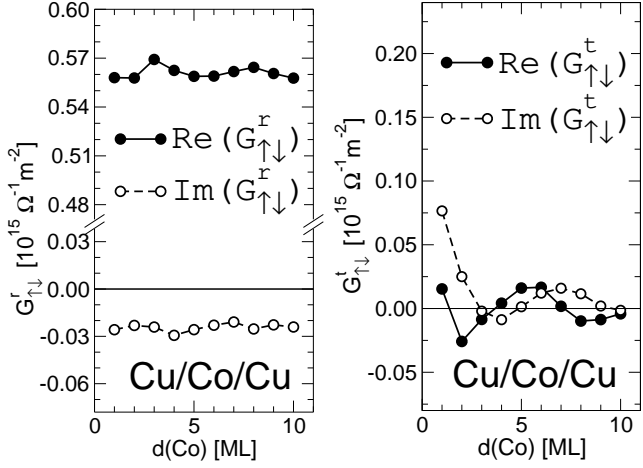

\includegraphics[height=0.26\textheight,clip=]{gmix_dis_cuco.eps} %
\includegraphics[height=0.26\textheight,clip=]{tmix_dis_cuco.eps}
\caption{Spin-mixing conductances of a Cu/Co/Cu (111) trilayer with
disordered interfaces as a function of the thickness $d$ of the Co layer.}
\label{gccd}
\end{figure}

Figures \ref{gafd} and \ref{gccd} show the same quantities
($G_{\uparrow \downarrow }^{r}$ and $G_{\uparrow \downarrow }^{t}$)
calculated in the presence of disorder modeled by 1 monolayer of
50~\% alloy added on each side of the magnetic layer. For both systems
we have used $10 \times 10$ lateral supercells.  The thickness $d$ in this case
is that of the clean ferromagnetic layer. For both material systems,
the effect of disorder is to strongly reduce the amplitudes of the
oscillations. The reflection mixing conductance becomes
practically
constant at the level of its asymptotic (i.e. interfacial) value. For $%
G_{\uparrow \downarrow }^{t}$, the oscillations are not entirely damped out
but their amplitude is substantially reduced. In fact, the values of $%
G_{\uparrow \downarrow }^{t}$ become negligible compared to $\mathrm{Re}%
G_{\uparrow \downarrow }^{r}$ for all but the thinnest magnetic layers. In
addition, we expect that diffusive scattering in the bulk of the magnetic
layer, which for simplicity has not been included here, will have a similar
effect.

In view of the above results, we conclude that in a typical situation $%
A^{\uparrow \downarrow }\approx g_{\uparrow \downarrow }^{r}$, where $%
g_{\uparrow \downarrow }^{r}$ can be calculated simply for an \textit{N}/%
\textit{F} interface instead of a complete structure. The results of such
calculations are listed in Table~\ref{table} for clean and disordered
interfaces. The disorder here was modeled by 2~ML of 50~\% alloy. In spite
of this difference, the values are practically identical to the asymptotic
ones seen in Figs. \ref{gafc}, \ref{gccc}, \ref{gafd}, and \ref{gccd}. In
particular, $\mathrm{Im}G_{\uparrow \downarrow }^{r}$ assumes values two
orders of magnitude smaller than $\mathrm{Re}G_{\uparrow \downarrow }^{r}$,
with the latter being close to the Sharvin conductance of the normal metal.
This approximate equality results once again from a combination of amplitude
(small $|r^{\uparrow }|$) and uncorrelated spin-up and spin-down phase
effects.

The values given in Table~\ref{table} differ somewhat from ones
reported previously in Ref.~\onlinecite{Xia:prb02}. There are two
reasons for this.  Firstly, the calculations in
Ref.~\onlinecite{Xia:prb02} were performed using energy-independent
muffin-tin orbitals linearized about the centers of gravity of the
occupied conduction states. The current
implementation\cite{Zwierzycki:prb03,Xia:tbp} uses energy-dependent,
(non-linearized) MTO's, calculated exactly at the Fermi energy which
improves the accuracy of the method. Secondly, on performing the 2D-BZ
integration in Eq.~(\ref{eq:mix_cond}), it was assumed in Ref.~%
\onlinecite{Xia:prb02}\ that the contribution to the sum of
$\vec{k}_{||}$ points for which there are no propagating states in the
ferromagnet should be neglected. However, the lack of propagating
states in the ferromagnet does not necessarily prohibit the transfer
of spin angular momentum which can be mediated by evanescent states,
for example in the case of a magnetic insulator. The contribution from
such $\vec{k}_{||}$ points \textit{should} be included in the 2D-BZ
integration.

\begin{table}[t]
\begin{center}
\begin{tabular}{ccccccccc}
\hline\hline
System & Interface & $G^{\uparrow}$ & $G^{\downarrow}$ & $\mathrm{Re}
G^r_{\uparrow\downarrow}$ & $\mathrm{Im}G^r_{\uparrow\downarrow}$ & $G^{%
\text{Sh}}_{N}$ & $G^{\text{Sh}}_{F\uparrow}$ & $G^{\text{Sh}}_{F\downarrow}$
\\ \hline
Au/Fe & clean & 0.40 & 0.08 & 0.466 & 0.005 & 0.46 & 0.83 & 0.46 \\
(001) & alloy & 0.39 & 0.18 & 0.462 & 0.003 &  &  &  \\
Cu/Co & clean & 0.42 & 0.38 & 0.546 & 0.015 & 0.58 & 0.46 & 1.08 \\
(111) & alloy & 0.42 & 0.33 & 0.564 & -0.042 &  &  &  \\ \hline\hline
\end{tabular}%
\end{center}
\caption{Interface conductances in units of $10^{15}~\Omega^{-1}$m$^{-2}$.}
\label{table}
\end{table}

\subsection*{Comparison with experiment}

In Ref.~\onlinecite{Urban:prl01}, Urban \emph{et al.} reported
room-temperature (RT) observations of increased Gilbert damping for a
system consisting of two Fe layers separated by a Au spacer layer. The
magnetization of the thinner of the two ferromagnetic layers precesses
in the external magnetic field. The other ferromagnetic layer,
with the direction of its magnetization fixed, acts as a spin sink.
No modification of the damping coefficient was measured for
configurations without a second Fe layer.
The latter finding is consistent with the prediction given
by Eq.~\eqref{tA} in the  $\lambda _{\text{SD}}\to \infty$ limit (well fulfilled for Au) as
discussed in Sec.~\ref{sec:theory}.

\begin{figure}[tbp]
\includegraphics[width=0.365\textheight,clip=]{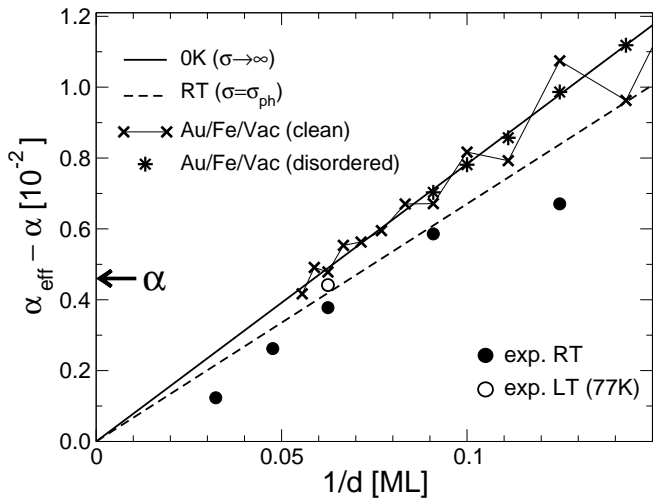}
\caption{Enhancement of the Gilbert damping coefficient for an Fe/Au/Fe
trilayer as a function of $1/d$ where $d$ is the thickness of the excited
Fe layer. The filled circles ($\bullet $) are the RT values measured in
Ref.~\onlinecite{Urban:prl01} and the open one ($\circ $) is a
low temperature value from Ref.~\onlinecite{Heinrich}.
The theoretical predictions based on Eq.~(\protect\ref{gilbert_coeff})
for 0~K (with $\protect\sigma \to \infty $) are shown as solid
and the RT-corrected (with phonon scattering) ones as dashed lines.
The results of 0~K calculations for a Au/Fe/vacuum system are given by
crosses  (\textbf{\textsf{x}})
and stars ($\ast$) for specular and disordered interfaces,
respectively. The value of the Gilbert damping $\alpha$ for a single Fe film
is marked with an arrow.}
\label{fig:urban}
\end{figure}

In the presence of a second Fe layer, Eq.~\eqref{Af} should be used.
Neglecting $\mathrm{Im}\tilde{A}^{\uparrow\downarrow}_{F\mathrm{/}N\mathrm{/}F}$ leads to
$\gamma_{\text{eff}}=\gamma$ and the damping enhancement
\begin{equation}
\alpha _{\mathrm{eff}}-\alpha=\frac{\hbar \gamma \mathrm{Re}\tilde{A}_{F%
\mathrm{/}N\mathrm{/}F}^{\uparrow \downarrow }}{4\pi M_{s}d}\,,
\label{gilbert_coeff}
\end{equation}%
where $\alpha\approx0.0046$ is the damping measured for a single
layer.  Using\cite{Heinrich:prl87} $\gamma=2.1\mu _{B}/\hbar $ and the
values of the interface and Sharvin conductances from
Table~\ref{table} (we assume that the values are the same for both
Au/Fe interfaces), Eq.~(\ref{gilbert_coeff}) is compared with the
experimental data\cite{note5} in Fig.~\ref{fig:urban} for various
assumptions about $\sigma$ in (\ref{Af}).  In the low temperature
limit and neglecting the residual resistivity of the Au layer,
$\sigma\to \infty $, Eq.~\eqref{gilbert_coeff} yields the solid line
which is seen to overestimate the damping enhancement compared to the
measured results. Using finite values of $\sigma$ will lead to lower
values of $\tilde{A}^{\uparrow\downarrow}$ and indeed, it was found
experimentally\cite{Heinrich} that lowering the temperature
(increasing the conductivity) increases the damping by as much as
about 20~\% (open circle in Fig.~\ref{fig:urban}).  If we use the room
temperature (RT) conductivity due to phonon scattering in crystalline
bulk Au,\cite{webelements} $\sigma_{\text{ph}} =0.45\times
10^{8}~\Omega ^{-1}\mathrm{m}^{-1}$, the dashed line is obtained
which, as expected, is closer to the RT measurements. The agreement with
experiment can be further improved by taking into account the possibility
of non-negligible residual resistance\cite{note3} of the Au spacer.
Assuming, for example, $\sigma_{\text{res}} =0.45\times
10^{8}~\Omega ^{-1}\mathrm{m}^{-1}$ and $\sigma_{\text{ph}} = 0$
would obviously yield the dashed line in the figure while taking
$\sigma_{\text{res}} = \sigma_{\text{ph}} = 0.45\times
10^{8}~\Omega ^{-1}\mathrm{m}^{-1} $ and
$1/\sigma=1/\sigma_{\text{res}}+1/\sigma_{\text{ph}}$ will yield
a line very close to the measured points.


The theoretical results represented by the straight lines in
Fig.~\ref{fig:urban} are based upon the asymptotic, single-interface
value of $G_{\uparrow \downarrow }^{r}$ from Table~\ref{table},
assuming $G_{\uparrow \downarrow }^{t}$ to be zero. To study possible
size-dependent corrections in thin films, the experimental system
needs to be represented by a more realistic model than the symmetric
\textit{N{\rm/}F{\rm/}N} structures discussed in the previous section.
The Au/Fe/GaAs structure used in Ref.~\onlinecite{Urban:prl01} differs
from these in two important respects.
First, the transmission mixing conductance ($g^t_{\uparrow\downarrow}$)
is identically zero because of the insulating substrate.
Secondly, because the reflection is perfect for {\em both} spin channels,
the thickness-dependent terms in Eq.~(\ref{eq:msc_refl}) have larger
amplitudes, leading to more pronounced oscillations of
$g^r_{\uparrow\downarrow}$ than those seen in Figs.~\ref{gafc} and
\ref{gccc}.  To estimate the variation which
can result from size-dependent corrections, we have performed a series
of calculations for a Au/Fe/vacuum structure, using vacuum instead of
GaAs for simplicity. The mixing conductance for the other Au/Fe
interface is kept at its asymptotic value (Table \ref{table}).
The results for perfect (specular)
structures,\cite{note4} marked in Fig.~\ref{fig:urban} with black
crosses (\textbf{\textsf{x}}), exhibit oscillations of non-negligible
amplitude about the asymptotic values given by the solid line
(arbitrarily taking the low-temperature regime, i.e. $\sigma\to\infty$
for reference).  The introduction of interface disorder (two
ML of 50~\%-50~\% alloy) yields values for the damping [stars ($\ast$)
in Fig.~\ref{fig:urban}] essentially averaged back to the limit given
by the single-interface calculations of Table~\ref{table}.

We have thus demonstrated that direct first-principles calculations
can produce values of the damping coefficient in the same range as
those measured experimentally. What is more, by taking into account
various other sources of scattering in the Au spacer and/or quantum-size
effects, the calculations can be brought into very close
agreement with experiment. A more definitive quantitative comparison
with experiment would require a detailed knowledge of the
microscopic structure of the experimental system which is currently
not available.
\subsection*{Spin current induced torque}
The mixing conductances calculated above, which describe how a spin current
flows through the system in response to an externally applied spin
accumulation $\pmb{\mu}$ (defined as a vector with length equal to
half of the spin-splitting of the chemical potentials
$|\pmb{\mu}|=(\mu_\uparrow-\mu_\downarrow )/2$), also describe the spin torque exerted on the
moment of the magnetic layer (see \textit{e.g}. Refs.
\onlinecite{Slon96,Brataas:prl00,Stiles:prb02,Xia:prb02,Brataas:spin_echo}).
Consider for example the situation where the spin
accumulation has been induced by some means in the left lead only and the
ferromagnet is magnetized along the $z$ axis. The spin
current incident on the interface is proportional to the number of incoming
channels in the lead $\mathbf{I}_{\mathrm{in}}^{\mathrm{L}}=\frac{1}{2\pi}g_{\mathrm{N%
}}^{\mathrm{Sh}}\pmb{\mu}$ whereas the transmitted spin current is given by%
\cite{Brataas:spin_echo}
\begin{equation}
\mathbf{I}_{\mathrm{out}}^{\mathrm{R}}=\frac{1}{2\pi}\left(
\begin{array}{ccc}
\mathrm{Re}g_{\uparrow \downarrow }^{\mathrm{t}} & \mathrm{Im}g_{\uparrow
\downarrow }^{\mathrm{t}} & 0 \\
-\mathrm{Im}g_{\uparrow \downarrow }^{\mathrm{t}} & \mathrm{Re}g_{\uparrow
\downarrow }^{\mathrm{t}} & 0 \\
0 & 0 & (g^{\uparrow }+g^{\downarrow })/2%
\end{array}%
\right) \pmb{\mu}  \label{sc_trans}
\end{equation}%
and the reflected spin current by%
\begin{equation}
\mathbf{I}_{\mathrm{out}}^{\mathrm{L}}=\frac{1}{2\pi}\left(
\begin{array}{ccc}
g_{\mathrm{N}}^{\mathrm{Sh}}-\mathrm{Re}g_{\uparrow \downarrow }^{\mathrm{r}} &
-\mathrm{Im}g_{\uparrow \downarrow }^{\mathrm{r}} & 0 \\
\mathrm{Im}g_{\uparrow \downarrow }^{\mathrm{r}} &
g_{\mathrm{N}}^{\mathrm{Sh}} - \mathrm{Re}g_{\uparrow\downarrow }^{\mathrm{r}} & 0 \\
0 & 0 & g_\mathrm{N}^\mathrm{Sh}-(g^{\uparrow }+g^{\downarrow })/2 %
\end{array}%
\right) \pmb{\mu}  \label{sc_reflect}
\end{equation}%
where $g^{\sigma }=\sum_{nm}|t_{nm}^{\sigma }|^{2}$ are the conventional
Landauer-B\"{u}ttiker conductances. Therefore $g_{\uparrow
\downarrow }^{\mathrm{t}}$ determines the transverse
component of the transmitted spin current subject to precession and absorption
within the magnetic layer.
Similarly, the real and imaginary parts of $g_{\mathrm{N}}^{%
\mathrm{Sh}}-g_{\uparrow \downarrow }^{r}=\sum_{mn}r_{mn}^{\uparrow
}r_{mn}^{\downarrow \star }$ are related to
the components of the reflected
transverse spin current. The rapid decay of $g_{\uparrow \downarrow
}^{t}$ (and $g_{\mathrm{N}}^{\mathrm{Sh}}-g_{\uparrow \downarrow }^{r}$) discussed
in previous paragraphs as a function of increasing magnetic layer thickness
implies that the absorption of the transverse component of the spin current
occurs within a few monolayers of the \textit{N{\rm/}F} interface.
In particular we find that the presence of the disorder improves
the effectiveness of the absorption.
The limit $g_{\uparrow \downarrow }^{t}\to 0$ and $g_{\uparrow
\downarrow }^{r}\to g_{\mathrm{N}}^{\mathrm{Sh}}$ corresponds to the
situation where all of the incoming transverse polarized spin current is
absorbed in the magnetic layer. The torque is then proportional to the
Sharvin conductance of the normal metal. As demonstrated in Figs.~\ref{gafc}-%
\ref{tccc}, \ref{gafd}, and \ref{gccd}, this is the situation for all but
the thinnest (few monolayers) and cleanest magnetic layers.

\begin{table}[b]
\begin{center}
\begin{ruledtabular}
\begin{tabular}{ccccc}
                                        &  Cu   & Ta    & Pd   &  Pt   \\    \hline
$D(\varepsilon_F)$(states/Ry-atom-spin) &  2    & 10    & 15   &  12   \\
$[1 - D(\varepsilon_F) I_{xc}]^{-1}$    &  1.1  & 1.9   & 4.4  &  2.2  \\
$G^{Sh} (10^{15}~\Omega^{-1}$m$^{-2}$)  &  0.56 & 0.97  & 0.62 &  0.68 \\
\end{tabular}%
\end{ruledtabular}
\end{center}
\caption{Density of states at the Fermi level, Stoner enhanement
factor and typical Sharvin conductances for
bulk {\em fcc} Cu, Pd and Pt and {\em bcc} Ta. Typical values
of the Stoner parameter, $I_{xc}$, were taken from references
\onlinecite{Gunnarsson:jpf76} and \onlinecite{Janak:prb77}.}
\label{tableIII}
\end{table}

\subsection*{Material Dependence}

The input parameters of spin pumping theory are scattering matrix
elements which are computed using the effective potential of Kohn-Sham
theory. This potential is calculated self-consistently and includes
electron-electron interaction effects via an exchange-correlation
potential approximated using the local spin density approximation,
and the Hartree potential.
In particular, the modification of interface parameters as a result of
magnetic moments being induced in the normal metal by proximity to a
ferromagnet (discussed in the Appendix of Ref.\onlinecite{Simanek:prep})
is already included in our results in a self-consistent and
non-perturbative manner (see Table \ref{tableI}). For the Cu and Au
normal metals we have considered, this effect is small. Expressed in
terms of a Stoner enhancement, this is related to the low Fermi level
densities of states of these metals, $D(\varepsilon_F)$. Viewing it in
this way poses the question of the possibility of finding much larger
effects for materials such as Pd and Pt which have a large density of
states at the Fermi level (see Table~\ref{tableIII}) and are known to be
close to a ferromagnetic transition as expressed by the susceptibility
enhancement $\chi / \chi_0 = [1 - D(\varepsilon_F) I_{xc}]^{-1}$, also
included in the Table. To calculate this factor, typical values of the
Stoner parameter, $I_{xc}$, were taken from references
\onlinecite{Gunnarsson:jpf76} and \onlinecite{Janak:prb77}.

To examine whether enhancements of the Gilbert damping parameter
recently reported \cite{Mizukami:jjap01,Mizukami:jmmm01} for thin
layers of Ta, Pd and Pt compared to Cu are related to their large
Fermi level densities of states, we need to reexamine how the
electronic structure enters our description of the Gilbert damping.
In the spin-pumping formulation, the quantities determining the
damping enhancement are not densities of states but transmission and
reflection mixing conductances determined from the scattering matrix.
These, we have seen, can be approximated very well by
$\mathrm{Re}G_{\uparrow \downarrow }^{r}$ which is very close to the
Sharvin conductance of the normal metal. Values of this quantity are
given for Cu, Ta, Pd and Pt in the last row of the Table. It is seen
that the Sharvin conductance changes less than $D(\varepsilon_F)$.
More significantly, with a maximum for Ta, the trend does not
correspond to that observed experimentally:
\cite{Mizukami:jjap01,Mizukami:jmmm01} Cu $\rightarrow$ Ta
$\rightarrow$ Pd $\rightarrow$ Pt. We believe that the explanation
should be sought elsewhere, possibly in the increasing spin-orbit
interaction which will lead to the heavier materials behaving as more
efficient spin sinks. To examine this suggestion in detail requires
the reformulation of the spin-pumping theory to include spin-orbit
interaction.

\section{Conclusions}

\label{sec:conclusions} In summary, we have calculated the transmission and
reflection mixing conductances that govern the non-local effects in the
ferromagnetic magnetization dynamics for two commonly used \textit{N}/%
\textit{F} combinations: Au/Fe and Cu/Co. In both cases, the transmission
mixing conductance $g_{\uparrow \downarrow }^{t}$ is much smaller than the
reflection mixing conductance $g_{\uparrow \downarrow }^{r}$, except for the
thinnest magnetic films, only a few atoms thick. Even for such thin films, $%
g_{\uparrow \downarrow }^{t}$ is smaller than $g_{\uparrow \downarrow}^{r}$.
Furthermore, $g_{\uparrow \downarrow }^{t}$ is more sensitive to disorder,
even a small amount of which reduces it to zero while having only a small
effect on $g_{\uparrow \downarrow }^{r}$ as shown in Figs.~\ref{gafd},\ref%
{gccd}. For all thicknesses, $\mathrm{Re}g_{\uparrow \downarrow }^{r}\gg
\mathrm{Im}g_{\uparrow \downarrow }^{r}$ and $\mathrm{Re}g_{\uparrow
\downarrow }^{r}$ is very close to its interfacial value (i.e. the
mixing conductance of the infinitely-thick magnetic film). The general
formulas (\ref{geff}) and (\ref{aeff}) predict that the spin pumping
renormalizes both the Gilbert damping ($\alpha $) and the gyromagnetic ratio
($\gamma $) of a ferromagnetic film embedded in a conducting non-magnetic
medium. However, in view of the results discussed in the previous section,
we conclude that, for all but the thinnest and cleanest magnetic layers, the
only effect of the spin pumping is to enhance the Gilbert damping. The
correction is directly proportional to the real part of the reflection
mixing conductance and is essentially an interface property. We also find
that oscillatory effects are averaged out for realistic band structures,
especially in the presence of disorder. $\mathrm{Re}g_{\uparrow \downarrow
}^{r}$ (which determines the damping enhancement of a single ferromagnetic
film embedded in a perfect spin-sink medium) is usually very close to $g_{%
\mathrm{N}}^{\text{Sh}}$ for intermetallic interfaces\cite%
{Xia:prb02,Stiles:prb02} (being in general bounded by $2g_{\mathrm{N}}^{%
\text{Sh}}$ according to its definition, Eq.~\ref{eq:mix_cond}).
These results also apply to the spin-current induced
magnetization reversal in intermetallic systems, indicating that the
\textquotedblleft effective field\textquotedblright\ correction due to the
imaginary part of the mixing conductance and bulk contributions to the
torque are very small.

\begin{figure}[t]
\includegraphics[width=0.365\textheight,clip=on]{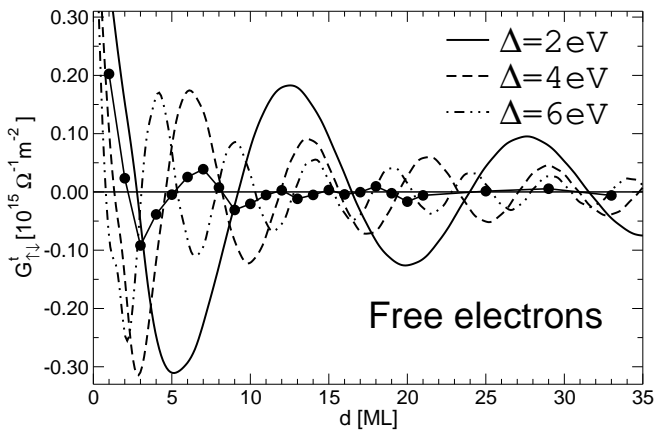}
\caption{The real part of $G^t_{\uparrow\downarrow}$
  calculated for a free electron model with
  $\varepsilon_F=7\;\mathrm{eV}$ (energy measured from the bottom of
  the parabolic conduction band in the normal metal) and various
  choices of the exchange splitting $\Delta$. The interlayer distance
  is taken to be the same as for Cu/Co(111) system.  The results of
  the first principles calculations ($\bullet$) from Fig.~\ref{tccc}
  are included for comparison.}
\label{tfree}
\end{figure}

\begin{acknowledgments}
This work is part of the research program for the \textquotedblleft
Stichting voor Fundamenteel Onderzoek der Materie\textquotedblright\ (FOM)
and the use of supercomputer facilities was sponsored by the
\textquotedblleft Stichting Nationale Computer
Faciliteiten\textquotedblright\ (NCF), both financially supported by the
\textquotedblleft Nederlandse Organisatie voor Wetenschappelijk
Onderzoek\textquotedblright\ (NWO). It was also supported by the European
Commission's RT Network \emph{Computational Magnetoelectronics} (Contract
No. HPRN-CT-2000-00143), by the NEDO International Joint Research Grant
Program \emph{Nano-Magnetoelectronics} and by The Harvard Society of
Fellows. MZ wishes also to acknowledge support from KBN grant
No.~PBZ-KBN-044/P03-2001.
\end{acknowledgments}

\appendix

\section{Comparison with a free electron model}
A combination of interfacial and bulk dephasing mechanisms, discussed
in Sec.~\ref{sec:results-discussion}, ensures that in the asymptotic
(thick magnetic layer) limit the spin-pumping mixing-conductance,
$A^{\uparrow\downarrow}$, reduces to the reflection mixing conductance
$G^r_{\uparrow\downarrow}$, with the latter quantity assuming values
which are predominantly real and equal to those determined for a
single N/F interface. Thinner layers exhibit oscillatory behaviour
which is most pronounced for $G^t_{\uparrow\downarrow}$
(Figs.~\ref{tafc} and \ref{tccc}). The amplitude of oscillation
however is at most 20\% of the asymptotic value of
$G^r_{\uparrow\downarrow}$ and decreases to less than 5\% for layers
more than 10ML thick.  This fast decay, found even for clean, fully
coherent structures, contrasts with results reported in
Ref.~\onlinecite{Mills:prb03} for a free-electron model.
For thin layers, Mills found the damping coefficient oscillated
with amplitude in the range of 80\% of the asymptotic value and,
for layers several tens of MLs thick, it was still of order 10\%.
This feature of the free-electron model is illustrated vividly in
Fig.~\ref{tfree} by comparing Re$(G^t_{\uparrow\downarrow})$ for
Cu/Co/Cu(111) from Fig.~4 with the corresponding results calculated
for free-electrons. In our free-electron calculation, the Fermi energy
in the non-magnetic material was taken to be 7eV in order to obtain
the correct value for the Sharvin conductance of Cu and the effect of
changing the exchange splitting $\Delta$ of the ferromagnet was
studied.  For $\Delta = 2, 4, 6$~eV, the amplitude of oscillation is
much larger and the decay is much slower than what we find for the
more realistic multi-band electronic structures. As might be expected,
increasing the exchange-splitting from 2 to 6~eV leads to a shorter
period and more rapid decay of the oscillations. However, in order to
mimic the parameter-free result, an exchange splitting in the range of
10~eV would be needed (not shown in the figure). Such a large value
cannot be justified either on theoretical or experimental grounds.
This discrepancy illustrates the difficulty of mapping the complex
electronic structure of transition metals onto single band models in a
meaningful way. Free-electron models do not adequately describe the
effectiveness of the thickness-dependent ``bulk'' dephasing in the
ferromagnet.  What is more, they also cannot reproduce the complex
spin- and $\vec{k}_{||} $-dependence of the interface scattering
coefficients (illustrated in Fig.~\ref{fig:2d_plots}) which results
from the mismatch of the normal metal electronic structure and the
quite different majority- and minority-spin electronic structures of a
ferromagnetic metal.  For single band free-electron models, the
interface scattering coefficients contain much less structure and
consequently this model fails to take into account even qualitatively
the dephasing effect of the interface.



\begin{thebibliography}
\bibitem{note1} Note that there is a sign error in
  Refs.~\onlinecite{Tserkovnyak:prl021} and \onlinecite{Tserkovnyak:prb022}:
  the quantity $A_i$ defined in Eq. (6) of Ref.~\onlinecite{Tserkovnyak:prl021}
  is equal to $-\mathrm{Im}A^{\uparrow\downarrow}$,
  not $\mathrm{Im}A^{\uparrow\downarrow}$.
\bibitem{note2} The renormalization of the
  mixing conductances has a close analogy in the
  discussion of various ways of measuring the resistance in
  mesoscopic systems.
  It is argued (see, e.g. Ref.~\onlinecite{Datta:book})
  that for the standard conductances, the ``bare'' quantities
  $g=\sum_{mn}|t_{mn}|^2$ are suitable for the description of
  two-point measurements with the reservoirs acting as both the
  current and voltage contacts. For an idealized four-point
  measurements with the voltage drop being measured directly across
  the scatterer (e.g. an interface, as in our case) the renormalized
  conductance (with the contact resistance subtracted) is
  needed. In the simplest limit of a single-mode
  conductor, with $g^{\rm Sh}=1$, the renormalized conductance reduces to
  the ``old'' Landauer formula $\tilde{g}=g/(1-g)$.
  It is intuitively clear that these are also the quantities that one
  needs to use when the scatterer is embedded in the diffuse environment.
\bibitem{note3} Measurements of the sheet conductivity\cite{Heinrich}
  indicate that the Au layers used in the experiments may have
  non-negligible residual resistances. The conductivity entering
  Eq.~(\ref{Af}) should not include interfacial-scattering
  contributions so the measurement of Ref. \onlinecite{Heinrich}
  cannot be used directly to obtain $\sigma_{\text{res}}$.
\bibitem{note4} It should be stressed, however, that although this
  calculation might give an estimate of the quantum-size effects, the
  applicability of Eq.~(\ref{Af}) requires that the system be globally
  diffuse.\cite{Bauer:prb03} The treatment of
  Ref.~\onlinecite{Bauer:prb03} would need to be generalized to enable
  consideration of nearly specular structures with little bulk disorder.
\end{thebibliography}
\end{document}